\documentclass[reprint,preprintnumbers,amsmath,amssymb,superscriptaddress,nofootinbib,preprintnumbers,floatfix,aps,prc]{revtex4-2}

\usepackage{graphicx}
\usepackage{dcolumn}

\usepackage{bigdelim}
\usepackage{amsmath}
\usepackage{amssymb}
\usepackage{lineno}
\usepackage{isotope}
\usepackage[unicode]{hyperref}
\usepackage{multirow}
\usepackage{xspace}

\usepackage[position=bottom, caption=false]{subfig}

\newcommand{\phantomsubfloat}[1]{
    {
        \captionsetup[subfigure]{labelformat=empty}
        \subfloat[][]{#1}
    }%
}

\hypersetup{
    colorlinks=true, 
    linkcolor=blue,  
   citecolor=red
}

\begin{document}

\title{Determination of the titanium spectral function from \texorpdfstring{$(e,e^\prime p)$}{(e,e'p)} data}

\author{L. Jiang} \affiliation{Center for Neutrino Physics, Virginia Tech, Blacksburg, Virginia 24061, USA}
\author{A.~M.~Ankowski} \affiliation{SLAC National Accelerator Laboratory, Stanford University, Menlo Park, California 94025, USA}
\author{D.~Abrams} \affiliation{Department of Physics, University of Virginia, Charlottesville, Virginia 22904, USA}
\author{L.~Gu} \affiliation{Center for Neutrino Physics, Virginia Tech, Blacksburg, Virginia 24061, USA}
\author{B.~Aljawrneh} \affiliation{North Carolina Agricultural and Technical State University, Greensboro, North Carolina 27401, USA}
\author{S.~Alsalmi} \affiliation{King Saud University, Riyadh 11451, Kingdom of Saudi Arabia}
\author{J.~Bane} \affiliation{The University of Tennessee, Knoxville, Tennessee 37996, USA}
\author{A.~Batz} \affiliation{The College of William and Mary, Williamsburg, Virginia 23187, USA}
\author{S.~Barcus} \affiliation{The College of William and Mary, Williamsburg, Virginia 23187, USA}
\author{M.~Barroso} \affiliation{Georgia Institute of Technology, Georgia 30332, USA}
\author{V.~Bellini} \affiliation{INFN, Sezione di Catania, Catania, 95123, Italy}
\author{O.~Benhar} \affiliation{INFN and Dipartimento di Fisica, Sapienza Universit\`{a} di Roma, I-00185 Roma, Italy}
\author{J.~Bericic} \affiliation{Thomas Jefferson National Accelerator Facility, Newport News, Virginia 23606, USA}
\author{D.~Biswas} \affiliation{Hampton University, Hampton, Virginia 23669, USA}
\author{A.~Camsonne} \affiliation{Thomas Jefferson National Accelerator Facility, Newport News, Virginia 23606, USA}
\author{J.~Castellanos} \affiliation{Florida International University, Miami, Florida 33181, USA}
\author{J.-P.~Chen} \affiliation{Thomas Jefferson National Accelerator Facility, Newport News, Virginia 23606, USA}
\author{M.~E.~Christy} \affiliation{Thomas Jefferson National Accelerator Facility, Newport News, Virginia 23606, USA}
\author{K.~Craycraft} \affiliation{The University of Tennessee, Knoxville, Tennessee 37996, USA}
\author{R.~Cruz-Torres} \affiliation{Massachusetts Institute of Technology, Cambridge, Massachusetts 02139, USA}
\author{H.~Dai} \affiliation{Center for Neutrino Physics, Virginia Tech, Blacksburg, Virginia 24061, USA}
\author{D.~Day} \affiliation{Department of Physics, University of Virginia, Charlottesville, Virginia 22904, USA}
\author{A.~Dirican} \affiliation{Department of Physics, University of Maryland, College Park, Maryland 20742, USA}
\author{S.-C.~Dusa} \affiliation{Thomas Jefferson National Accelerator Facility, Newport News, Virginia 23606, USA}
\author{E.~Fuchey} \affiliation{University of Connecticut, Storrs, Connecticut 06269, USA}
\author{T.~Gautam} \affiliation{Hampton University, Hampton, Virginia 23669, USA}
\author{C.~Giusti} \affiliation{Dipartimento di Fisica, Universit\`{a} degli Studi di Pavia and INFN, Sezione di Pavia,  I-27100 Pavia, Italy}
\author{J.~Gomez}\thanks{deceased} \affiliation{Thomas Jefferson National Accelerator Facility, Newport News, Virginia 23606, USA}
\author{C.~Gu} \affiliation{Duke University, Durham, North Carolina 27708, USA}
\author{T.~J.~Hague} \affiliation{Kent State University, Kent, Ohio 44242, USA}
\author{J.-O.~Hansen} \affiliation{Thomas Jefferson National Accelerator Facility, Newport News, Virginia 23606, USA}
\author{F.~Hauenstein} \affiliation{Old Dominion University, Norfolk, Virginia 23529, USA}
\author{D.~W.~Higinbotham} \affiliation{Thomas Jefferson National Accelerator Facility, Newport News, Virginia 23606, USA}
\author{C.~Hyde} \affiliation{Old Dominion University, Norfolk, Virginia 23529, USA}
\author{Z.~Jerzyk} \affiliation{Department of Physics, St. Norbert College, De Pere, Wisconsin 54115, USA}
\author{A.~M.~Johnson} \affiliation{Department of Physics, Gettysburg College, Gettysburg, Pennsylvania 17325, USA}
\author{C.~Keppel} \affiliation{Thomas Jefferson National Accelerator Facility, Newport News, Virginia 23606, USA}
\author{C.~Lanham} \affiliation{Center for Neutrino Physics, Virginia Tech, Blacksburg, Virginia 24061, USA}
\author{S.~Li} \affiliation{University of New Hampshire, Durham, New Hampshire 03824, USA}
\author{R.~Lindgren} \affiliation{Department of Physics, University of Virginia, Charlottesville, Virginia 22904, USA}
\author{H.~Liu} \affiliation{Columbia University, New York, New York 10027, USA}
\author{C.~Mariani}\email{mariani@vt.edu} \affiliation{Center for Neutrino Physics, Virginia Tech, Blacksburg, Virginia 24061, USA}
\author{R.~E.~McClellan} \affiliation{Thomas Jefferson National Accelerator Facility, Newport News, Virginia 23606, USA}
\author{D.~Meekins} \affiliation{Thomas Jefferson National Accelerator Facility, Newport News, Virginia 23606, USA}
\author{R.~Michaels} \affiliation{Thomas Jefferson National Accelerator Facility, Newport News, Virginia 23606, USA}
\author{M.~Mihovilovic} \affiliation{Jozef Stefan Institute, Ljubljana 1000, Slovenia}
\author{M.~Murphy} \affiliation{Center for Neutrino Physics, Virginia Tech, Blacksburg, Virginia 24061, USA}
\author{D.~Nguyen} \affiliation{Department of Physics, University of Virginia, Charlottesville, Virginia 22904, USA}
\author{M.~Nycz} \affiliation{Kent State University, Kent, Ohio 44242, USA}
\author{L.~Ou} \affiliation{Massachusetts Institute of Technology, Cambridge, Massachusetts 02139, USA}
\author{B.~Pandey} \affiliation{Hampton University, Hampton, Virginia 23669, USA}
\author{V.~Pandey} \altaffiliation{Present Address: Department of Physics, University of Florida, Gainesville, FL 32611, USA}
\affiliation{Center for Neutrino Physics, Virginia Tech, Blacksburg, Virginia 24061, USA}
\author{K.~Park} \affiliation{Thomas Jefferson National Accelerator Facility, Newport News, Virginia 23606, USA}
\author{G.~Perera} \affiliation{Department of Physics, University of Virginia, Charlottesville, Virginia 22904, USA}
\author{A.~J.~R.~Puckett} \affiliation{University of Connecticut, Storrs, Connecticut 06269, USA}
\author{S.~N.~Santiesteban} \affiliation{University of New Hampshire, Durham, New Hampshire 03824, USA}
\author{S.~\v{S}irca} \affiliation{Faculty of Mathematics and Physics, University of Ljubljana, Ljubljana 1000, Slovenia} \affiliation{Jozef Stefan Institute, Ljubljana 1000, Slovenia}
\author{T.~Su} \affiliation{Kent State University, Kent, Ohio 44242, USA}
\author{L.~Tang} \affiliation{Thomas Jefferson National Accelerator Facility, Newport News, Virginia 23606, USA} \affiliation{Hampton University, Hampton, Virginia 23669, USA}
\author{Y.~Tian} \affiliation{Shandong University, Shandong, 250000, China}
\author{N.~Ton} \affiliation{Department of Physics, University of Virginia, Charlottesville, Virginia 22904, USA}
\author{B.~Wojtsekhowski} \affiliation{Thomas Jefferson National Accelerator Facility, Newport News, Virginia 23606, USA}
\author{S.~Wood} \affiliation{Thomas Jefferson National Accelerator Facility, Newport News, Virginia 23606, USA}
\author{Z.~Ye} \affiliation{Physics Division, Argonne National Laboratory, Argonne, Illinois 60439, USA}
\author{J.~Zhang} \affiliation{Department of Physics, University of Virginia, Charlottesville, Virginia 22904, USA}

\collaboration{The Jefferson Lab Hall A Collaboration}

\begin{abstract}

The E12-14-012 experiment, performed in Jefferson Lab Hall A, has measured the $(e, e^\prime p)$ cross section in parallel kinematics using a natural titanium target. In this letter we report the analysis of the data-set obtained in different kinematics for our solid natural titanium target. Data were obtained in a range of missing momentum and missing energy respectively between $15 \lesssim p_m \lesssim 250$ MeV/c and $12 \lesssim E_m \lesssim 80$ MeV and using electron beam energy of 2.2~GeV. We measured the reduced cross section with a $\sim$7\% accuracy as function of both missing momentum and missing energy. Our Monte Carlo simulation, including both a model spectral function and the effects of final state interactions, reproduces satisfactorily the data.

\end{abstract}

\preprint{JLAB-PHY-22-3726}

\maketitle

{\it Introduction:}
The recent measurement of the \isotope[40][18]{Ar}$(e,e^\prime p)$ cross section\textemdash performed by the E12-14-012 collaboration in Jefferson Lab Hall A\textemdash has enabled the first determination of the spectral function describing the joint energy-momentum distribution of protons in the target nucleus~\cite{Jiang:2022}.

The JLab experiment, while also providing valuable new information on single-nucleon dynamics in complex nuclei, was primarily meant to obtain the input needed to improve the interpretation of data collected by neutrino experiments using liquid argon detectors, thus reducing the systematic uncertainty of  neutrino energy reconstruction. 

In principle, the extension of the analysis based on nuclear spectral functions to both neutrino and 
antineutrino interactions would require the availability of the neutron energy-momentum 
distribution in argon, whose experimental study using electron beams involves
challenging issues. 

The analysis of the data collected by a pioneering 
\isotope[4][]{He}$(e,e^\prime n)$ experiment, carried out at NIKHEF in the 1990s~\cite{4He:een},
has clearly demonstrated that\textemdash in contrast to the case of the $(e,e^\prime p)$ reaction\textemdash 
neutron knockout involves additional difficulties, associated with both the detection of the outgoing 
neutron and a reliable identification of the reaction mechanism. 
A comparison between the results of theoretical calculations and the measured cross section, corresponding to 
momentum transfer $q = 250$~MeV$/c$ and missing momentum in the range $25 < p_{\rm miss} < 70$ MeV$/c$,
shows that in this kinematic regime charge-exchange processes\textemdash in which the detected neutron 
is {\it not} produced at the elementary interaction vertex\textemdash provide the dominant contribution, and must be carefully taken into account.

An alternative, admittedly rather crude, procedure to obtain information on 
the neutron distribution in argon is based on the observation that the neutron spectrum of \isotope[40][18]{Ar} is mirrored by the proton spectrum of the nucleus of titanium, having charge $Z=22$. Based on this property, which reflects the isospin symmetry of nuclear forces, it has been argued that the proton spectral function obtained from {Ti}$(e,e^\prime p)$ data provides a viable proxy for the neutron spectral function of argon~\cite{Dai:2018xhi}. The validity of this hypothesis is supported by the results of Ref.~\cite{Barbieri:2019ual}, whose authors have 
employed the proton and neutron spectral functions of argon obtained from a state-of-the-art theoretical model to 
carry out an accurate calculation of the double-differential \isotope[40][18]{Ar}$(\nu_\mu,\mu^-)$ cross section. The results obtained replacing the neutron spectral function of argon with the proton spectral function of titanium turn out to be in remarkably good agreement; in fact, they are nearly indistinguishable from one another.  

In this letter, we report the results of the analysis of the {Ti}$(e,e^\prime p)$ data collected in Jefferson Lab Hall A by the E12-14-012 collaboration, and discuss the representation of the reduced cross sections in terms of a model proton spectral function.

{\it Experimental setup:}
%
Experiment E12-14-012 was approved by the Jefferson Lab PAC in 2014 and data was taken in the Spring 2017. In the past few years a series of measurements have been completed: the inclusive, $(e,e^\prime)$~\cite{Dai:2018xhi,Dai:2018gch,Murphy:2019wed}, and exclusive $(e,e^\prime p)~$\cite{Gu:2020rcp,Jiang:2022} electron scattering cross sections on several targets, including a natural gas argon target~\cite{Gu:2020rcp,Jiang:2022}. 

An electron beam of 2.2~GeV and $\approx 22~\mu$A was provided by the Jefferson Lab Continuous Electron Beam Accelerator Facility (CEBAF). The scattered protons and electrons were detected in coincidence in two nearly identical high-resolution spectrometers (HRSs) both consisting of a dipole and three quadrupole magnets. The electron and proton spectrometers are both equipped with vertical drift chambers (VDCs)~\cite{Fissum:2001st}, scintillator planes (two) for timing measurements and triggering, and a double-layered lead-glass calorimeter. In addition, the electron arm is equipped with a gas \v{C}erenkov counter for particle identification and pion rejectors, while the proton arm is equipped with pre-shower and shower detectors~\cite{Alcorn:2004sb}. The experimental kinematics used during data taking on the natural titanium target were identical for kinematic 2-4 to  those used for the Ar target~\cite{Jiang:2022} and in the case of kinematic 1 the missing energy was set to 50~MeV.

\begin{figure}[tb]
    \centering
    \includegraphics[width=0.8\columnwidth]{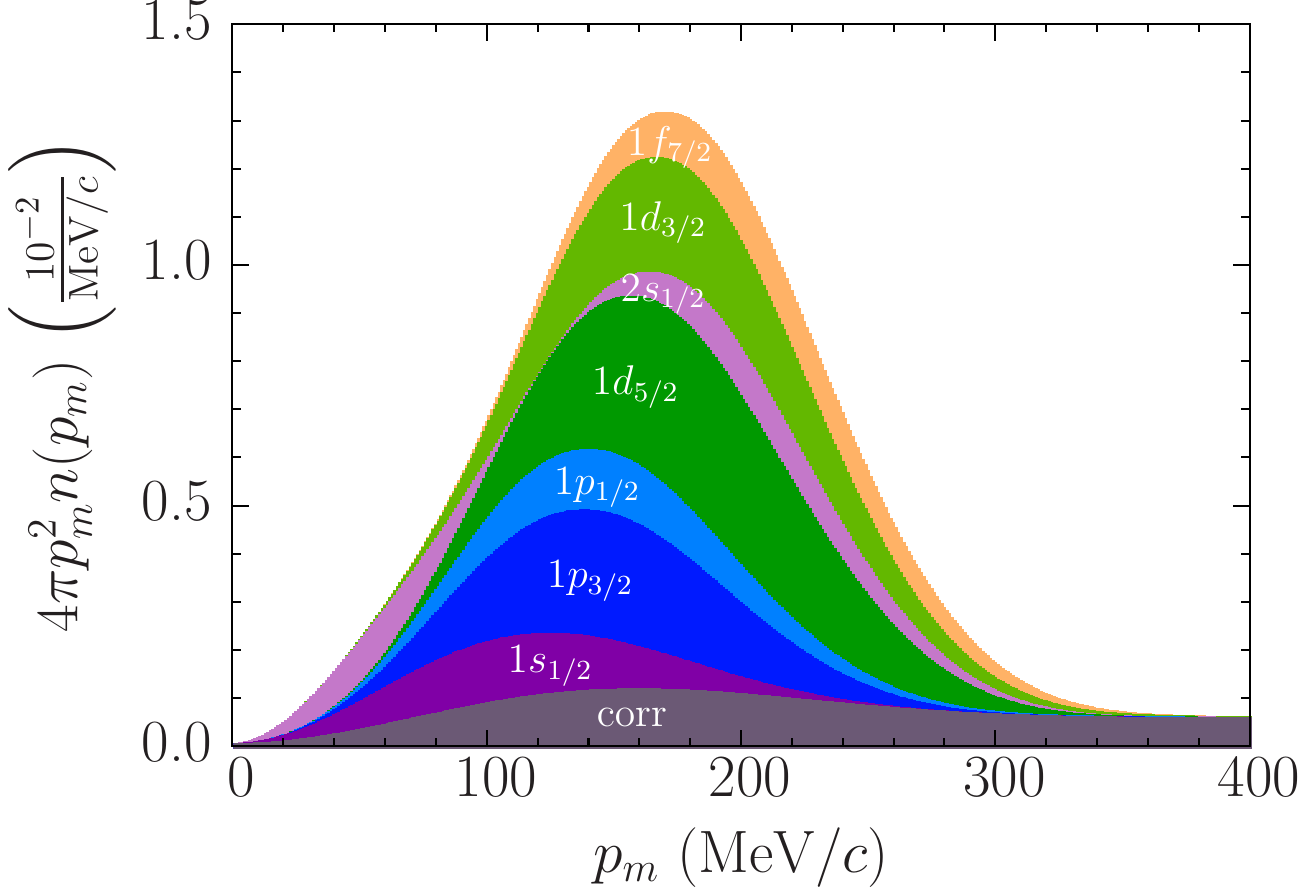}
    \caption{Missing momentum distribution of protons in titanium in the test spectral function, presented with the geometric factor of $4\pi p_m^2$.}
    \label{fig:missing_momentum_distribution}
\end{figure}

We determine the six-fold differential cross section as a function of $p_{m}$ and $E_{m}$ following the same method as described in Ref.~\cite{Jiang:2022}.
The reduced cross section was obtained as a function of $p_m$ and $E_m$, from the double differential cross section using the elementary electron-proton off-shell cross section $\sigma_{ep}$ of de~Forest~\cite{Dieperink:1976wy,DeForest:1983ahx}. The simulated momentum distributions are presented in Fig.~\ref{fig:missing_momentum_distribution}. 
The missing energy of the shell-model states is assumed to follow the Gaussian distribution as in Ref.~\cite{Jiang:2022}, with the peak positions determined as described in detail elsewhere~\cite{Gu:2020rcp}.

The JLab SIMC spectrometer package~\cite{SIMC} was used to simulate $(e,e^\prime p)$ events including an approximate spectral function for Ti, geometric details of the target, radiative corrections, and Coulomb effects.

\begin{table}[tb]
    \centering
    \caption{\label{tab:ti_energy_levels}Parametrization of the test spectral function of protons in titanium. The missing energy distribution of a~shell-model state $\alpha$ peaked at $E_{\alpha}$ has width $\sigma_{\alpha}$, and is normalized to the spectroscopic factor $S_\alpha$. For comparison, we also show the occupation number in the independent particle shell-model $N_\alpha$. For the correlated part, we provide the total normalization and the threshold for two-nucleon knockout $E_\text{thr}$.}
    \begin{ruledtabular}
    \begin{tabular}{ c d d d d }
     $\alpha$  & \multicolumn{1}{c}{$N_\alpha$} & \multicolumn{1}{c}{$S_\alpha$} & \multicolumn{1}{c}{$E_\alpha$ (MeV)} & \multicolumn{1}{c}{$\sigma_\alpha$ (MeV)} \\
     \colrule
       $1f_{7/2}$ & 2 & 1.6 & 11.45 & 2 \\
       $1d_{3/2}$ & 4 & 3.2 & 12.21 & 2 \\
       $2s_{1/2}$ & 2 & 1.6 & 12.84 & 2 \\
       $1d_{5/2}$ & 6 & 4.8 & 15.46 & 4 \\
       $1p_{1/2}$ & 2 & 1.6 & 35.0 & 6  \\
       $1p_{3/2}$ & 4 & 3.2 & 40.0 & 6 \\
       $1s_{1/2}$ & 2 & 1.6 & 62.0 & 10 \\
       corr. & \text{---} & 4.4 & 22.09 & \text{---} \\
    \end{tabular}
    \end{ruledtabular}
\end{table}

The correlated spectral function is estimated within the approach of Ref.~\cite{CiofidegliAtti:1995qe}, as a convolution
integral involving the momentum distributions of the
relative and center-of-mass motion of a correlated proton-neutron pair~\cite{Jiang:2022}. By construction, the correlated part accounts for 20\% of the total strength of the test spectral function, see Table~\ref{tab:ti_energy_levels}. Compared with the correlated spectral function of argon, the one for titanium differs due to different mass number $A$, and the employed value of the $pn$ knockout threshold. These differences translate into a~higher absolute number of correlated $pn$ pairs in titanium than in argon, and slightly different energy of the residual system. On the other hand, the parameters of the relative motion of the $pn$ pairs and their center-of-mass motion  are assumed not to differ between argon and titanium.

\begin{figure}[bt]
    \centering
    \phantomsubfloat{\label{fig:SFa}}
    \phantomsubfloat{\label{fig:SFb}}
    \includegraphics[width=0.8\columnwidth]{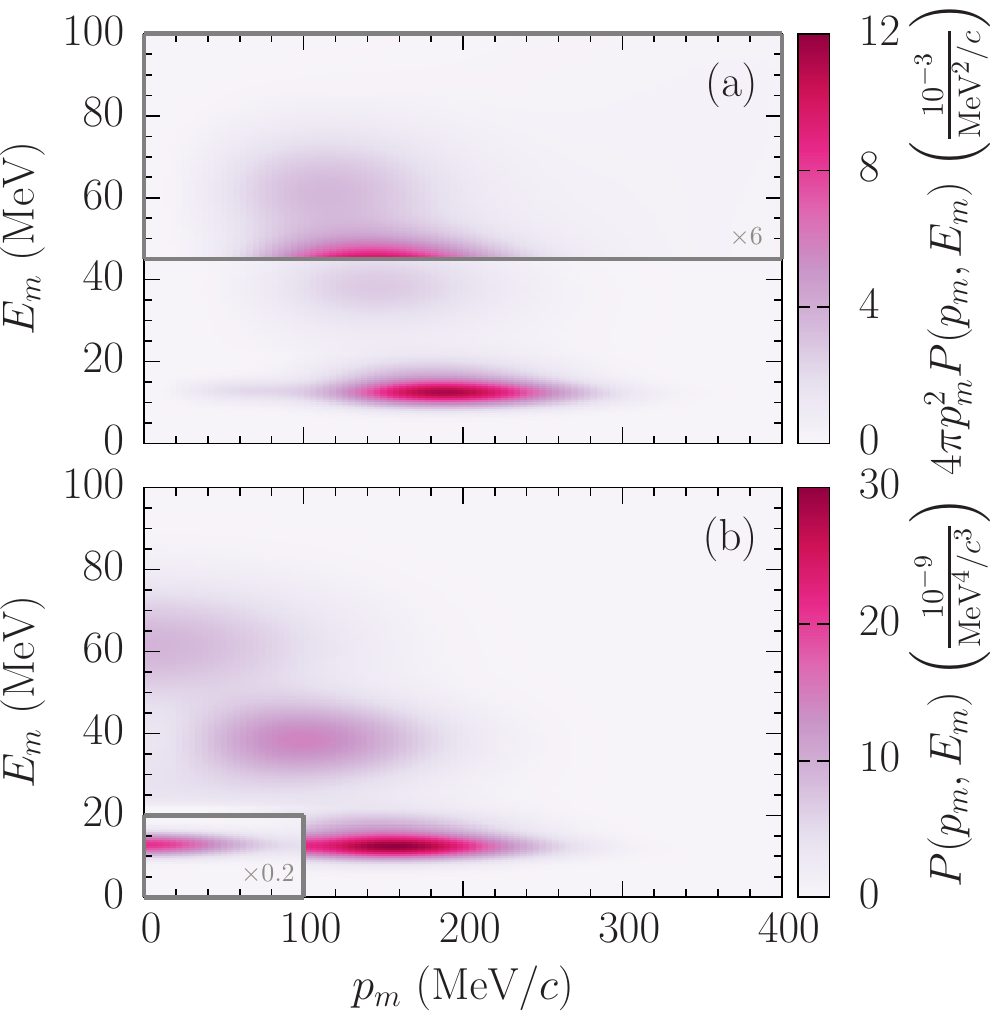}
    \caption{Test spectral function shown (a) with and (b) without the geometric factor of $4\pi p_m^2$. Note that multiplicative factors are used for clearer presentation of some regions.}
\label{fig:SF}
\end{figure}

Figure~\ref{fig:SF} displays the test spectral function as a function of missing momentum and missing energy. 

We observe an energy resolution for the Ti states comparable to the one of the Ar target, this is due mainly to the FSI corrections in our MC simulation that contribute to a broadening of the states.

{\it Data Analysis:}
\begin{table}[tb!]
\caption{\label{tab:syst}Contributions to systematic uncertainties for titanium averaged over all the $E_m$ and $p_m$ bins for each kinematics. All numbers are in \%. For kin4, the results correspond to the systematic uncertainties of the signal and the background added in quadrature.}
\begin{ruledtabular}
\begin{tabular}{@{}l c c c c c @{}}
					&						            	& kin1     		& kin2		& kin3		& kin4	   	\\
\hline
{1.~Total statistical uncertainty} &			        				& 0.78   		& 0.60		& 0.82		& 1.24		\\
{2.~Total systematic uncertainty}  &			        				& 4.63	   		& 4.92		& 4.70		& 6.04		\\
\phantom{1. }a.~Beam $x\&y$ and HRS offset &				& 0.75			& 1.71		& 1.19		& 1.47  	\\
\phantom{1. }c.~Optics (q1, q2, q3) &						& 0.48			& 0.77		& 0.55		& 0.90		\\
\phantom{1. }d.~Acceptance cut $(\theta,\phi, z)$ & 		& 1.36			& 1.46		& 1.32		& 1.57		\\
\phantom{1. }e.~Target thickness/density/length & 			& 0.20			& 0.20		& 0.20		& 0.20	    \\
\phantom{1. }f.~Calorimeter \& \v{C}erenkov \& $\beta$ cuts& & 0.29			& 0.58		& 0.42		& 2.83		\\
\phantom{1. }g.~Radiative and Coulomb corr. & 				& 1.00			& 1.00		& 1.00		& 1.00		\\
\phantom{1. }i.~Cross section model and FSI &				& 4.12			& 2.23		& 2.23		& 2.23		\\
\phantom{1. }j.~Trigger and coincidence time cut &			& 0.78			& 0.33		& 0.58		& 2.32		\\

\end{tabular}
\end{ruledtabular}
\end{table}
The total systematic uncertainty in this analysis is the sum in quadrature of the individual uncertainties as listed in Table~\ref{tab:syst}. We followed the same procedure as described in Ref.~\cite{Jiang:2022}: kinematic and acceptance cuts are considered uncorrelated bin to bin and they do not depend on the theoretical input model. All the kinematic and acceptance cuts were varied according to the variable's resolution. The simulation did not contain a correction for final state interaction (FSI) effects other than the transparency corrections.

We repeated the analysis of systematic uncertainties varying all MC input parameters, and the effect on the analysis was found to be negligible. To determine the uncertainties related to the target position, we varied the simulation's inputs within uncertainties, and we used a different optical transport matrix obtained using independent variation of each of the three quadrupole magnetic fields. For each of this variation we computed the effect with respect to a run where no variations were used and the differences were summed in quadrature. The total systematic uncertainty is then computed using the same assumptions and recipe as described in Ref.~\cite{Gu:2020rcp,Jiang:2022}.

We set our $E_m$ ranges, 0 to 30 MeV, 30 to 54 MeV, and 54 to 90 MeV. We then integrated the missing energy distribution for each of the ranges and performed the fit on the corresponding missing momentum distribution to improve our sensitivity.

We  fit the experimental missing energy and missing momentum distributions to extract spectroscopic factors, mean value and width of each of the \isotope[48][]{Ti} orbitals.

For each bin in the spectra of missing energy (100 $\times$ 1~MeV) and missing momentum (40 bins with different range in momentum depending upon the kinematics), we computed the reduced MC cross section~\cite{Arrington:1998ps} and the ratio of the data to simulation yield, and we combined them as the following:
\begin{equation}
 \frac{d^{2}\sigma^\text{red}_{cc1}}{d\Omega dE'} = \Big(\frac{d^{2}\sigma^\text{red}_{cc1}}{d\Omega dE'}\Big)_{\text{MC}} \times \frac{Y(E',\theta)}{Y_{\text{MC}}(E',\theta)},
 \label{eq:yield_for_chi2}
\end{equation}
where the $Y(E',\theta)$ is the yield for a given bin and the reduced MC cross section is a fit to the existing data~\cite{SIMC}. The reduced cross section includes (i) the $\sigma_{cc1}$ cross section of de~Forest~\cite{DeForest:1983ahx}, (ii) the predictions of the spectral function model, (iii) radiative corrections~\cite{Mo:1968cg},  (iv) Coulomb corrections~\cite{Aste:2005wc}, and  (v) FSI corrections, described within the distorted-wave impulse approximation (DWIA) scheme.

The fit performs a $\chi^2$ minimization using the {\sc minuit}~\cite{James:2004vla} package available in {\sc root}~\cite{ROOT}. 

\par The $\chi^2$ function is defined as:
\begin{equation}
\chi^2 = \sum_i\chi^2_i = \sum_i\left ( \frac{ \sigma_i^\text{red, obs} - \sum_\alpha{S_\alpha f_\alpha^\text{pred}(i)} }{\sigma_{\sigma_i^\text{red, obs}}} \right )^2,
\label{eq:chi2_i}
\end{equation}
where the index $i$ labels the missing momentum bin, $\alpha$ is the orbital index, $f_\alpha^\text{pred}(i)$ is the parametrized prediction evaluated at bin $i$ in the missing momentum spectra for orbital $\alpha$, $S_\alpha$ is the spectroscopic factor. 
The missing momentum distribution, as in the case of the Ar,
 does not show dependence on the mean energies and widths of the orbitals. 

Table~\ref{tab:momentum_fit_results} provides a summary of the fit of the missing momentum distributions including the degrees of freedom and the value of the $\chi^2$. The fit was repeated excluding the correlated part of the SF to avoid possible bias.

The spectroscopic factors reported in Tables~\ref{tab:momentum_fit_results} and \ref{tab:momentum_energy_spectra} are normalized to $80\% \times 22$ for the total strength of the orbitals and to $20\% \times 22$ for the correlated part and they include corrections due to phase space coverage.

\begin{table}
\centering
\caption{\label{tab:momentum_fit_results} Comparison of the results of the $\chi^2$ minimization using the missing momentum distributions, determined with and without the use of the correlated spectral function. For every state $\alpha$, we determined the spectroscopic factor $S_\alpha$, and its occupation number in an independent-particle shell model, $N_\alpha$. We include the total spectroscopic strength, the number of degrees of freedom (d.o.f.), and the $\chi^2$ per d.o.f.}
\begin{ruledtabular}
\begin{tabular}{c c r r }
   & &  \multicolumn{1}{c}{w/ corr.}      & \multicolumn{1}{c}{w/o corr.} \\
\colrule
$\alpha$    & $N_\alpha$   & \multicolumn{2}{c}{$S_\alpha$}  	\\
\colrule
$1f_{7/2}$  & 2         & $0.83\pm1.17$ & \phantom{0}$0.78\pm1.35$ \\
$1d_{3/2}$  & 4         & $1.17\pm0.22$ & \phantom{0}$1.34\pm0.10$   \\
$2s_{1/2}$  & 2         & $2.02\pm0.08$	& \phantom{0}$2.18\pm0.08$  \\
$1d_{5/2}$  & 6	        & $2.34\pm1.34$	& \phantom{0}$2.34\pm3.72$  \\
$1p_{1/2}$  & 2	        & $2.46\pm0.27$	& \phantom{0}$2.71\pm1.19$  \\
$1p_{3/2}$  & 4	        & $5.46\pm1.69$	& \phantom{0}$5.46\pm0.05$  \\
$1s_{1/2}$  & 2	        & $2.17\pm0.09$	& \phantom{0}$2.51\pm0.08$  \\
corr.       & 0         & $5.15\pm0.41$ & \multicolumn{1}{c}{excluded}\\
\colrule
$\sum_\alpha S_\alpha$  &      & $21.60\pm2.51$          & $17.32\pm4.20$ \\
d.o.f. &   & 675         & 676  \\
$\chi^2$/d.o.f.&      & 0.49           &   0.57                     \\
\end{tabular}
\end{ruledtabular}
\end{table}

We then repeat the fit minimizing the $\chi^2$ function using the missing energy spectra,
\begin{equation}
\chi^2 = \sum_i{\chi^2_i} + \sum_n{\left (\frac{\tau_n^\textrm{fit} - \tau_n^c}{\sigma_n^\textrm{fit}}\right)^2}.
\label{eq:chi2_constraints}
\end{equation}
We include additional constraints (summarized in Tab.~\ref{tab:Constraints}) in the form of penalty terms to the $\chi^2$ function using data from Ref.~\cite{Chen:2018trb,Wang:2017,Mairle:1993asu,Mairle:1993ntp}.

\begin{table}[tb]
\centering
\caption{\label{tab:Constraints} External constraints on the fits to the missing-energy spectra computed using data from past measurements~\cite{Chen:2018trb,Wang:2017,Mairle:1993asu,Mairle:1993ntp}. For the clarity of presentation, we denote $E_\alpha$ as $E(\alpha)$.}
\begin{ruledtabular}
\begin{tabular}{c  d  d }
Parameter     		    & \multicolumn{1}{c}{Value (MeV)}  	    & \multicolumn{1}{c}{Uncertainty (MeV)} 	\\
\colrule
$E(1f_{7/2})$ & 11.32 & 0.10 \\
$E(1d_{3/2})$ & 12.30 & 0.24 \\
$E(2s_{1/2})$ & 12.77 & 0.25 \\
$E(1d_{5/2})$ & 15.86 & 0.20 \\
$E(1d_{5/2})-E(1d_{3/2})$  & 3.57  & 0.31 \\
$E(1p_{3/2})-E(1p_{1/2})$  & 6.36  & 0.75 \\
\end{tabular}
\end{ruledtabular}
\end{table}

\begin{table}
\centering
\caption{\label{tab:momentum_energy_spectra} Results of the $\chi^2$ minimization using the missing energy distributions for different cases. We repeated the fit with different priors, and not including the correlated part of the SF. For every state $\alpha$, we extract spectroscopic factor $S_\alpha$, occupational number $N_\alpha$ assuming independent particle model and total spectroscopic strength. We reported at the end also the number of degrees of freedom (d.o.f.), and the $\chi^2$ per d.o.f.}
\begin{ruledtabular}
\begin{tabular}{c c r r r } 
& & \multicolumn{1}{c}{all priors}     & \multicolumn{1}{c}{w/o $p_m$}      & \multicolumn{1}{c}{w/o corr.} \\
\colrule
$\alpha$    & $N_\alpha$    & \multicolumn{3}{c}{$S_\alpha$} \\      
\colrule
$1f_{7/2}$  & 2         & $1.53\pm0.25$ & $1.55\pm0.28$  & \phantom{0}$1.24\pm0.22$ \\
$1d_{3/2}$  & 4         & $2.79\pm0.37$ & $3.15\pm0.54$  & \phantom{0}$3.21\pm0.37$ \\
$2s_{1/2}$  & 2         & $2.00\pm0.11$	& $1.78\pm0.46$  & \phantom{0}$2.03\pm0.11$ \\
$1d_{5/2}$  & 6	        & $2.25\pm0.16$	& $2.34\pm0.19$  & \phantom{0}$3.57\pm0.29$ \\ 
$1p_{1/2}$  & 2	        & $2.00\pm0.20$	& $1.80\pm0.27$	 & \phantom{0}$2.09\pm0.19$ \\
$1p_{3/2}$  & 4	        & $2.90\pm0.20$	& $2.92\pm0.20$	 & \phantom{0}$4.07\pm0.15$ \\
$1s_{1/2}$  & 2	        & $2.14\pm0.10$	& $2.56\pm0.30$  & \phantom{0}$2.14\pm0.11$\\
corr.  	    & 0         & $4.71\pm0.31$ & $4.21\pm0.46$  & \multicolumn{1}{c}{excluded}	\\ 
\colrule
$\sum_\alpha S_\alpha$ & & $20.32\pm0.65$         & $20.30\pm1.03$           & $18.33\pm0.59$ \\
d.o.f       &           & 121           & 153            & 125 \\
$\chi^2$/d.o.f.&      & 0.95           & 0.71            & 1.23 \\
\end{tabular}
\end{ruledtabular}
\end{table}

The spin-orbit splitting has been computed using the phenomenological prescription of Ref.~\cite{Mairle:1993asu,Mairle:1993ntp},

$E(n,l,l-1/2)-E(n,l,l+1/2) = \frac{2 l + 1}{2n} k A^{-C},$
with angular momentum $l$, main quantum number $n$, and mass number $A$. The empirically determined constants $k=23.27$~MeV and $C=0.583$~\cite{Mairle:1993asu} are included in the fit as penalty function to the $\chi^2$. The uncertainty value has been calculated comparing the prediction of the  phenomenological prescription to the available experimental data from NIKHEF-K~\cite{Kramer:1989uiu,Kramer:1990,Leuschner:1994zz}. 

The fit on the missing energy spectra contains 23 parameters: 3 parameters for each orbital (the spectroscopic factor, the position of the maximum, and the width of the distribution) and 2 parameters for the correlated SF (the strength and the threshold energy).

We present our results in Table~\ref{tab:momentum_energy_spectra}. We repeated the fit excluding the results coming from the $p_m$ minimization and without the correlated SF part. 
All the results are compatible within errors, which indicates no large bias in the determination of the spectroscopic factors using a different set of constraints.

\begin{table}
\centering
\caption{\label{tab:energy_spectra_frpm_fits} Measured peak positions $E_\alpha$, widths $\sigma_\alpha$, and the parameter $E_\text{corr}$ of the correlated spectral function obtained from the fit of the missing energy distributions. Results are obtained including and excluding the results from the previous fit performed using the missing momentum.}
\begin{ruledtabular}
\begin{tabular}{c r r r r }
    & \multicolumn{2}{c}{$E_\alpha$ (MeV)}  	& \multicolumn{2}{c}{$\sigma_\alpha$ (MeV)} \\ 
\colrule
$\alpha$    & \multicolumn{1}{c}{w/ priors}     & \multicolumn{1}{c}{w/o priors}    & \multicolumn{1}{c}{w/ priors}        & \multicolumn{1}{c}{w/o priors} \\      
\colrule
$1f_{7/2}$  & $11.32\pm0.10$  & $11.31\pm0.10$    & \phantom{0}$8.00\pm5.57$      & \phantom{0}$8.00\pm6.50$ \\
$1d_{3/2}$  & $12.30\pm0.24$  & $12.33\pm0.24$    & \phantom{0}$7.00\pm0.61$    & \phantom{0}$7.00\pm3.84$ \\
$2s_{1/2}$ 	& $12.77\pm0.25$  & $12.76\pm0.25$    & \phantom{0}$7.00\pm3.76$    & \phantom{0}$7.00\pm3.84$ \\
$1d_{5/2}$	& $15.86\pm0.20$    & $15.91\pm0.22$    & \phantom{0}$2.17\pm0.27$    & $2.23\pm0.29$\\ $1p_{1/2}$	& $33.33\pm0.60$	& $33.15\pm0.65$	& $3.17\pm0.45$              & \phantom{0}$3.03\pm0.48$\\
$1p_{3/2}$	& $39.69\pm0.62$	& $39.43\pm0.68$	& \phantom{0}$5.52\pm0.70$    & \phantom{0}$5.59\pm0.70$ \\
$1s_{1/2}$	& $53.84\pm1.86$	& $52.00\pm3.13$ 	& $11.63\pm1.90$              & \phantom{0}$13.63\pm2.59$ \\
corr.  		& $25.20\pm0.02$    & $25.00\pm0.29$	    & \multicolumn{1}{c}{---}	& \multicolumn{1}{c}{---} \\
\end{tabular}
\end{ruledtabular}
\end{table}

\begin{figure}
    \begin{minipage}[h]{1.0\linewidth}
    \center{\includegraphics[width=1\linewidth]{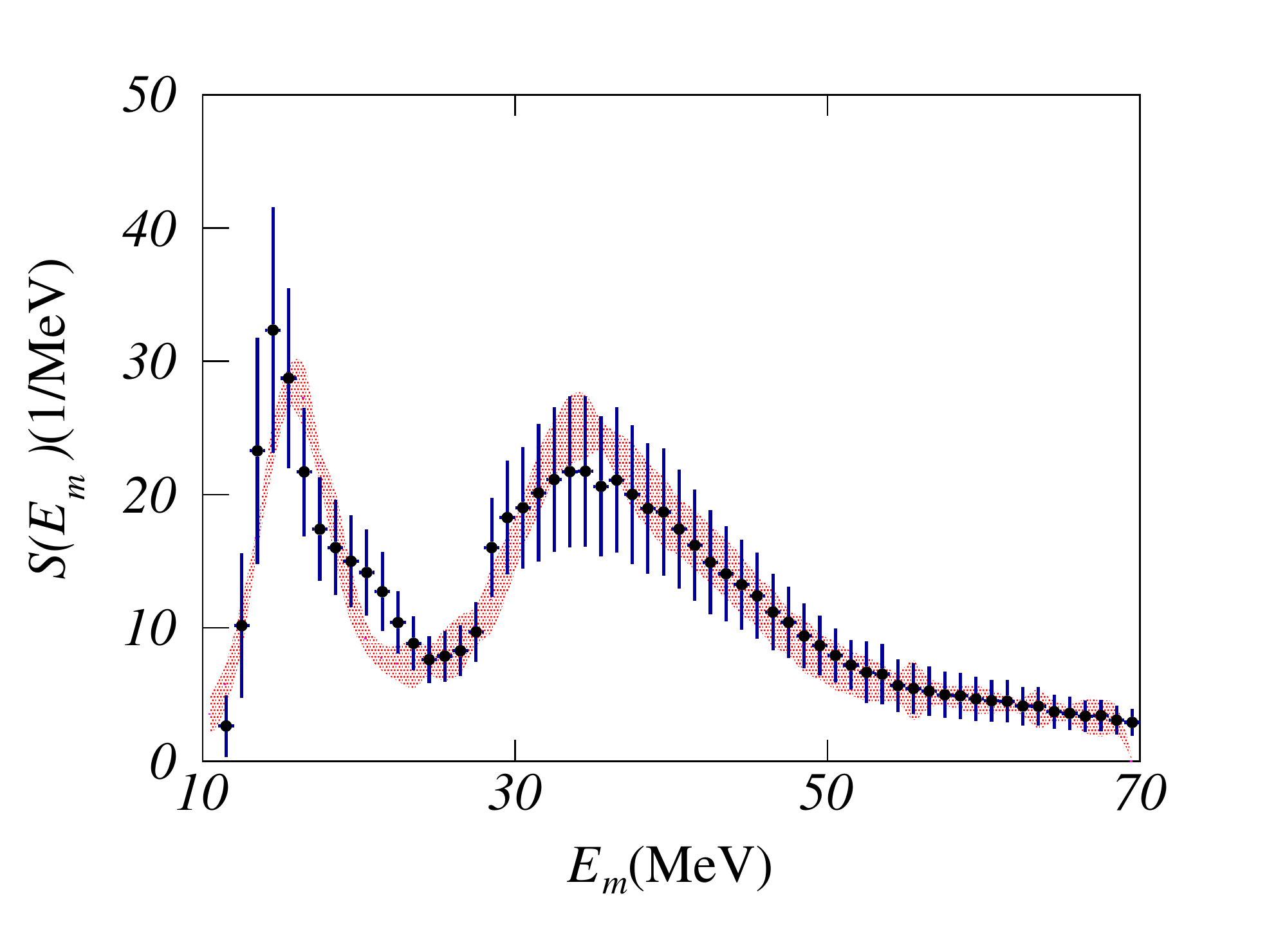}} 
    \end{minipage}
    \vfill
 \hfill
\caption{\label{fig:missing_energy_spectra1} Missing energy distributions obtained for natural titanium for $\quad130 <p_m < 260$~MeV/c. The red band indicates the final fit results including the full error uncertainties.}
\end{figure}

\begin{figure}[bt]
    \centering
\includegraphics[width=0.9\columnwidth]{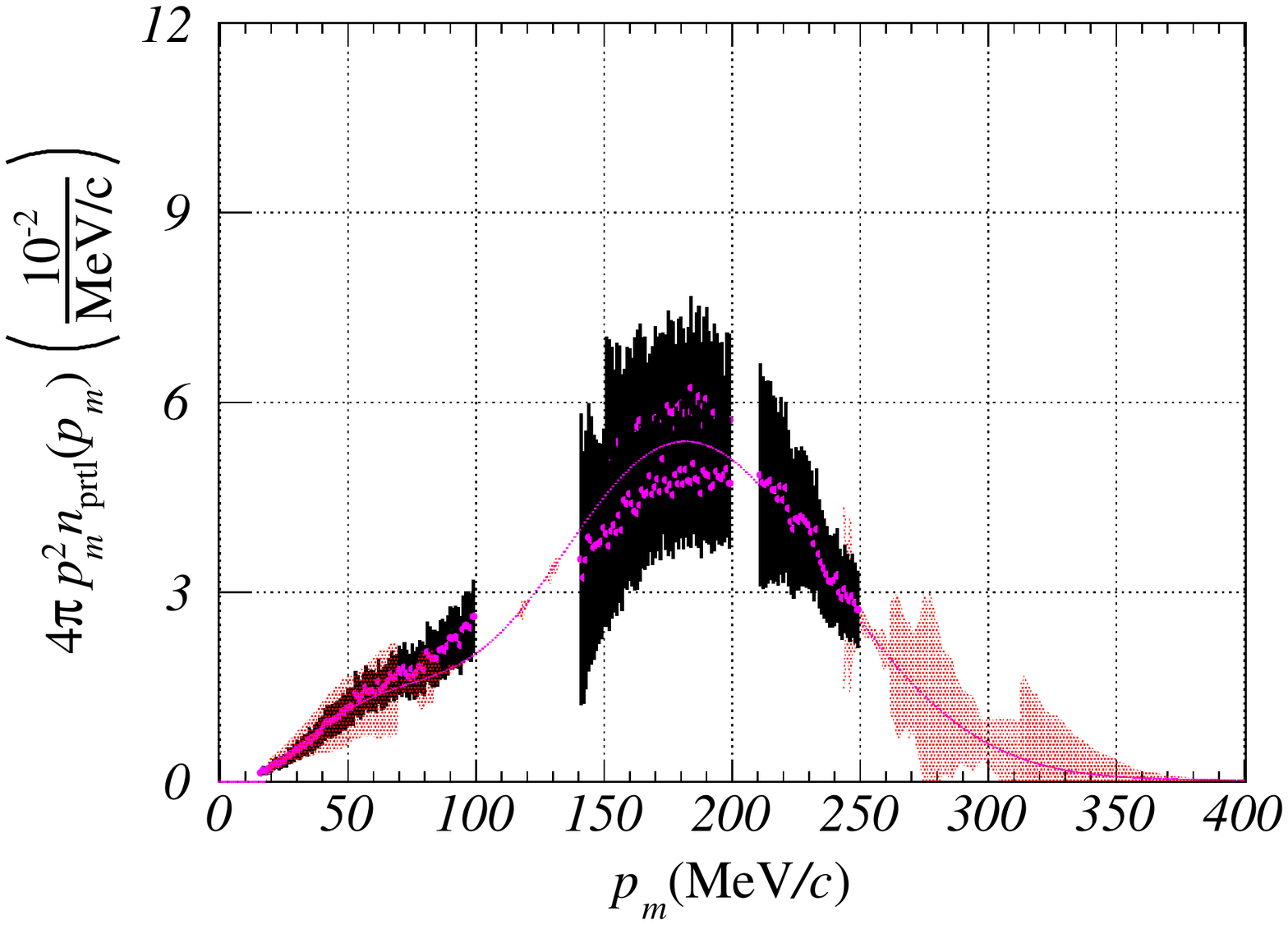}\qquad
       \caption{Partial momentum distribution obtained by integrating the test spectral function over the missing energy range of 10--30~MeV presented with the geometric factor of $4\pi p_m^2$. Different data regions represent data from different kinematics.}
\label{fig:missing_momentum_fit_results}
\end{figure}

As in our previous publications~\cite{Jiang:2022} we have evaluated the effects of different priors used in the fit including orbital modeling and parametrization and we found that the results are compatible within uncertainties.

Fig.~\ref{fig:missing_energy_spectra1} shows that the test spectral function model, rescaled using the parameters obtained from the fit, listed in Table~\ref{tab:momentum_energy_spectra}, is capable of reproducing our data satisfactorily.

Figure~\ref{fig:missing_momentum_fit_results} reports the missing momentum distribution obtained after integrating the data and the model over the missing energy range 30--54~MeV. 
It is apparent that the collected data cover the relevant kinematic range with just a few exceptions, most notably at vanishing $p_m$.
The experimental coverage is not complete due to experimental conditions and beam-time limitations. In particular, data for kinematics 4 is statistically limited. 

Fig.~\ref{fig:missing_momentum_fit_results} shows a good agreement\textemdash  within uncertainties\textemdash of the reduced cross sections using data from kinematics 2 and 3. As observed in Ref.~\cite{Jiang:2022}, this agreement supports, in Ti as in Ar, the validity of the factorisation scheme  used in computing the DWIA corrections.

{\it Summary and conclusions:}
The \isotope[][]{Ti}$(e,e^\prime p)$ data collected by experiment E12-14-012 at Jefferson Lab have been analysed to extract the proton spectral function in \isotope[][]{Ti}. The spectral function, an innate property of the nucleus, provides the energy and momentum distribution of protons bound in the titanium ground state.
Our spectral function derivation depends only from the model used to compute FSI corrections. The FSI uncertainties have been evaluated in the same way as in Ref.~\cite{Gu:2020rcp,Jiang:2022} and have been included in our measurements.

The results of our study of titanium provide important new and much needed information, and it will play a critical to further develop theoretical models capable of describing neutrino-nucleus interactions in liquid argon detectors like DUNE.

The results of the pioneering work of~Barbieri {\it et al.}~\cite{Barbieri:2019ual} that uses one of our previous results~\cite{Dai:2018xhi,Dai:2018gch,Murphy:2019wed} demonstrate the importance of the availability of electron scattering data in Ar and Ti. The work in Ref.~\cite{Barbieri:2019ual} showed that a replacement of the neutron SF of argon with the proton SF of \isotope[48][22]{Ti} in the calculation of the $\isotope[40][18]{Ar}(\nu_\mu,\mu^-)$ cross section at beam energy $E_\nu = 1$~GeV has a few-percent effect. It has to be kept in mind, however, that the inclusive cross section, which only involves integrals of the SFs, is rather insensitive to the details of the missing energy distributions. Therefore, the findings of Barbieri {\it et al.}, while being very encouraging, cannot be taken as clear-cut evidence of the validity of the assumption that the proton SF of natural titanium can be used as a proxy for the neutron SF of Ar, as suggested by isospin symmetry. More work will be necessary to put this hypothesis on a firm basis. Here we only note that our estimate of the top four energy levels of neutrons in \isotope[40][18]{Ar} suggests that they agree to 0.6--2.2 MeV with those of protons in \isotope[48][22]{Ti} listed in Table~\ref{tab:ti_energy_levels}.

The reduced differential cross sections has been fitted using a model 
spectral function. The effects of FSI, which are known to be significant in $(e,e^\prime p)$ reactions, have been included using the same factorization scheme which underlines our analysis and, as for the case of Ar~\cite{Jiang:2022}, seems to be reliable.

The comparison between data and MC simulation results has been shown in a broad range of missing energies, extending from the proton-knockout threshold
to $E_m \sim 80$ MeV. The overall agreement supports the validity of the theoretical basis of our analysis.    

We have determined the position and width of the peaks corresponding to shell model states, and estimated the corresponding spectroscopic strengths. 

A more accurate determination of the titanium spectral function will require a more advanced theoretical model of the energy and momentum distributions, as well as a refined implementation of the DWIA.

The extraction of the spectral function reported in this article\textemdash providing a satisfactory description of the proton energy and momentum distribution\textemdash should be seen as the achievement of the goals of the JLab experiment E12-14-012, and a step toward a more accurate description of (anti)neutrino interactions in argon.

The understanding of the proton and neutron spectral functions for argon will greatly improve the accuracy of neutrino and antineutrino energy reconstruction in measurements of neutrino oscillations, such as those in the short-baseline program of Fermilab and in the long-baseline studies in the Deep Underground Neutrino Experiment. 

As a final remark it should be pointed out that, up to FSI corrections, the factorisation ansatz\textemdash whose validity is clearly demonstrated by the observation of $y$-scaling ~\cite{Sick:1980ey}\textemdash provides the basis for the extraction of the spectral function from $(e,e^\prime p)$ data. 
The spectral function can be employed to describe initial state physics in any processes in which the beam particle couples to a single nucleon, including quasi elastic scattering, resonance production and deep inelastic scattering~\cite{Vagnoni:2017hll}. In correlated systems, these processes lead to the appearance of both 1p1h and 2p2h final states. On the other hand, the description of 2p2h final states originating from coupling to the two-nucleon meson-exchange currents requires an extension of the factorisation scheme, and the use of two-nucleon spectral functions, as discussed in Ref.~\cite{Benhar:2015ula}.


\begin{acknowledgments}
\par We acknowledge the outstanding support from the Jefferson Lab Hall A technical staff, target group and Accelerator Division. This experiment was made possible by Virginia Tech, the National Science Foundation under CAREER grant No. PHY$-$1352106 and grant No. PHY$-$1757087. This work was also supported by the DOE Office of Science, Office of Nuclear Physics, contract DE-AC05-06OR23177, under which Jefferson Science Associates, LLC operates JLab, DOE contract DE-FG02-96ER40950, DE-AC02-76SF00515, DE-SC0013615 and by the DOE Office of High Energy Physics, contract DE-SC0020262.
\end{acknowledgments}

%
%
%
%

\end{document}